\documentclass[aps, prd, floatfix, nofootinbib, superscriptaddress, twocolumn]{revtex4-1}

\usepackage{latexsym}
\usepackage{amsmath}
\usepackage{amssymb}
\usepackage{amsfonts}

\usepackage[mathscr,scaled=1.15]{urwchancal}
\DeclareFontFamily{OT1}{pzc}{}
\DeclareFontShape{OT1}{pzc}{m}{it}%
{<-> s * [1.15] pzcmi7t}{}
\DeclareMathAlphabet{\mathpzc}{OT1}{pzc}{m}{it}

\usepackage{color}

\usepackage{supertabular}
\usepackage{placeins}
\usepackage{epsfig}
\usepackage{graphicx}

\definecolor{purple}{rgb}{0.5,0,0.5}
\definecolor{blue}{rgb}{0.0,0,0.9}
\definecolor{prdblue}{rgb}{0.133,0.118,0.498}
\usepackage[colorlinks=true, pdfstartview=FitV, linkcolor=prdblue, citecolor= prdblue, urlcolor=prdblue]{hyperref}

\begin{document}


\title{$\,$\\[-6ex]\hspace*{\fill}{\normalsize{\sf\emph{Preprint no}.\ NJU-INP 016/20}}\\[1ex]
Semileptonic decays of $D_{(s)}$ mesons}

\author{Z.-Q.~Yao}
\affiliation{School of Physics, Nanjing University, Nanjing, Jiangsu 210093, China}
\affiliation{Institute for Nonperturbative Physics, Nanjing University, Nanjing, Jiangsu 210093, China}

\author{D.~Binosi}
\email[]{binosi@ectstar.eu}
\affiliation{European Centre for Theoretical Studies in Nuclear Physics
and Related Areas (ECT$^\ast$) and Fondazione Bruno Kessler\\ Villa Tambosi, Strada delle Tabarelle 286, I-38123 Villazzano (TN) Italy}

\author{Z.-F.~Cui}
\affiliation{School of Physics, Nanjing University, Nanjing, Jiangsu 210093, China}
\affiliation{Institute for Nonperturbative Physics, Nanjing University, Nanjing, Jiangsu 210093, China}

\author{C.\,D.~Roberts}
\email[]{cdroberts@nju.edu.cn}
\affiliation{School of Physics, Nanjing University, Nanjing, Jiangsu 210093, China}
\affiliation{Institute for Nonperturbative Physics, Nanjing University, Nanjing, Jiangsu 210093, China}

\author{S.-S.~Xu}
\affiliation{College of Science, Nanjing University of Posts and Telecommunications, Nanjing 210023, China}

\author{H.-S.~Zong}
\affiliation{School of Physics, Nanjing University, Nanjing, Jiangsu 210093, China}
\affiliation{Department of Physics, Anhui Normal, Nanjing University, Wuhu, 241000, China}
\affiliation{Nanjing Proton Source Research and Design Center, Nanjing 210093, China}

\date{08 March 2020}

\begin{abstract}
A symmetry-preserving continuum approach to meson bound-states in quantum field theory, employed elsewhere to describe numerous $\pi$- and $K$-meson electroweak processes, is used to analyse leptonic and semileptonic decays of $D_{(s)}$ mesons.
Each semileptonic transition is conventionally characterised by the value of the dominant form factor at $t=0$ and the following results are obtained herein:
$f_+^{D_s\to K}(0) = 0.673(40)$; $f_+^{D\to \pi}(0)=0.618(31)$; and $f_+^{D\to K}(0)=0.756(36)$.
Working with the computed $t$-dependence of these form factors and standard averaged values for $|V_{cd}|$, $|V_{cs}|$, one arrives at the following predictions for the associated branching fractions:
${\mathpzc B}_{D_s^+\to K^0 e^+ \nu_e} = 3.31(33)\times 10^{-3}$;
${\mathpzc B}_{D^0\to \pi^- e^+ \nu_e} = 2.73(22)\times 10^{-3}$;
and
${\mathpzc B}_{D^0\to K^- e^+ \nu_e} = 3.83(28)$\%.
Alternatively, using the calculated $t$-dependence, agreement with contemporary empirical results for these branching fractions requires $|V_{cd}|=0.221(9)$, $|V_{us}|=0.953(34)$.
With all $D_{(s)}$ transition form factors in hand, the nature of SU$(3)$-flavour symmetry-breaking in this array of processes can be analysed; and just as in the $\pi$-$K$ sector, the magnitude of such effects is found to be determined by the scales associated with emergent mass generation in the Standard Model, not those originating with the Higgs mechanism.
\vspace*{0ex}
\end{abstract}

\maketitle


\section{Introduction}
%
Working with a large sample of $e^+\,e^-$ collision data, acquired at the Beijing Electron Positron Collider (BEPC), the BESIII Detector Collaboration has released precise results on the semileptonic decays of $D_{(s)}$ mesons \cite{Ablikim:2015ixa, Ablikim:2018upe, Yuan:2019zfo}.  Combined with data from related analyses using the BaBar detector at the Stanford Linear Accelerator Center \cite{Lees:2014ihu}, the Belle detector in Japan \cite{Kou:2018nap} and the CLEO detector at Cornell University \cite{Besson:2009uv}, science now has a new window onto the Standard Model and beyond.  For example, these transition form factors can be used to provide increasingly tight constraints on the Cabbibo-Kobayashi-Maskawa matrix elements $|V_{cd}|$, $|V_{cs}|$ \cite{Tanabashi:2018oca}.

Notably, too, given that there is a heavy+light meson in the initial state and a pseudo-Nambu-Goldstone mode in the final state, sound theoretical analyses of such decays should provide fresh ways of revealing the interplay between explicit (Higgs-related) mass generation and emergent hadronic mass (EHM) in the Standard Model.  Insights here have the potential to expose facets of confinement dynamics.

Owing to their importance, the semileptonic decays of $D_{(s)}$ mesons have long been the subject of theoretical interest.  For instance, numerous phenomenological analyses have been completed, most recently Refs.\,\cite{Ivanov:2019nqd, Faustov:2019mqr}.  Also, the $D\to (\pi,K)$ transition form factors have been computed using lattice-regularised QCD (lQCD) \cite{Lubicz:2017syv}.  Notably, it is still necessary for lattice analyses to correct for discretisation-induced symmetry violations and employ extrapolations in order to reach physical light-quark current-masses \cite{Aoki:2019cca}.  Thus, comparison with results obtained using continuum Schwinger function methods (CSMs) can be valuable in both validating the lQCD results and enabling their insightful interpretation.

Herein, motivated by such considerations, we use the leading-order truncation of those equations required to complete a symmetry-preserving formulation of the continuum bound-state equations and deliver predictions for $D\to (\pi,K) \ell \nu_\ell$, $D_s \to K \ell \nu_\ell$ transition form factors.  In addition to complementing available lQCD computations, our results should: prove useful in constraining $|V_{cd}|$, $|V_{cs}|$; and expose the response of meson structure to the transition between the heavy-quark domain, within which the Higgs-mechanism dominates quark masses, and the light-quark sector, wherein EHM defines the characteristics of pseudoscalar mesons.  To ensure the study's reliability, we use existing continuum calculations of $\pi$- and $K$-meson leptonic and semileptonic decays as benchmarks \cite{Ji:2001pj, Chen:2012txa}.  Namely, current-quark masses are varied smoothly from those associated with $\pi$ and $K$ initial states up to those characterising $D_{(s)}$ mesons.  In doing so, we complete a unified description of the leptonic and semileptonic decays of the following systems: $\pi$, $K$, $D$, $D_s$.

The manuscript is arranged as follows.  Section~\ref{Sec2} describes the necessary transition matrix elements and our approximations to them.  The computational framework and associated algorithms are explained in Sec.\,\ref{SecCMR}, augmented by a collection of detailed appendices; and the results, their interpretation, and the insights they provide are canvassed in Sec.\,\ref{SecCI}.  Section~\ref{Epilogue} presents a summary and perspective.

\section{Transition Form Factors}
\label{Sec2}
\subsection{Observations on kinematics}
We consider the following matrix elements:
\begin{subequations}
\label{EqMEs}
\begin{align}
_d M_\mu^{D^+_s}(P,Q) &=\langle K^0(p) | \bar d i\gamma_\mu c |D_s^+(k)\rangle \nonumber \\
& = P_\mu f_+^{D_s^d}(t) + Q_\mu f_-^{D_s^d}(t)\,,\\
_d  M_\mu^{D^+}(P,Q) &=\langle \pi^-(p) | \bar d i\gamma_\mu c |D^0(k)\rangle  \nonumber \\
& = [P_\mu f_+^{D_u^d}(t) + Q_\mu f_-^{D_u^d}(t)]\,,\\
_s  M_\mu^{D^+}(P,Q) &=\langle \bar K^0(p) | \bar s i\gamma_\mu c |D^+(k)\rangle \nonumber \\
& = P_\mu f_+^{D_d^s}(t) + Q_\mu f_-^{D_d^s}(t)\,,\\
_s  M_\mu^{D^0}(P,Q) &=\langle K^-(p) | \bar s i\gamma_\mu c |D^0(k)\rangle \nonumber \\
& = P_\mu f_+^{D_d^s}(t) + Q_\mu f_-^{D_d^s}(t)\,,
\end{align}
\end{subequations}
where the last line is true so long as isospin symmetry is assumed;
$P =k+p$,
$Q=p-k$,
with $k^2 = -m_{D_{(s)}}^2$ and $p^2=-m_{K,\pi}^2$, depending on the initial and final state;
and the squared-momentum-transfer is $t=-Q^2$.\footnote{In our Euclidean metric conventions:  $\{\gamma_\mu,\gamma_\nu\} = 2\delta_{\mu\nu}$; $\gamma_\mu^\dagger = \gamma_\mu$; $\gamma_5= \gamma_4\gamma_1\gamma_2\gamma_3$, tr$[\gamma_5\gamma_\mu\gamma_\nu\gamma_\rho\gamma_\sigma]=-4 \epsilon_{\mu\nu\rho\sigma}$; $\sigma_{\mu\nu}=(i/2)[\gamma_\mu,\gamma_\nu]$; $a \cdot b = \sum_{i=1}^4 a_i b_i$; and $Q_\mu$ timelike $\Rightarrow$ $Q^2<0$.}

Naturally, the masses of the hadrons involved limit the physically accessible range of the transition form factors:
\begin{subequations}
\begin{align}
P\cdot Q & = - (m_{D_{(s)}}^2 - m_{K,\pi}^2) =: - \Delta_{D_{(s)} (K,\pi)}\,,\\
P^2 & = - 2 (m_{D_{(s)}}^2 + m_{K,\pi}^2) - Q^2 =: -2 \Sigma_{D_{(s)} (K,\pi)} - Q^2;
\end{align}
\end{subequations}
and $t_m^{D_{(s)} (K,\pi)} = (m_{D_{(s)}} - m_{K,\pi})^2 =: m_{D_{(s)}}^2 y_m^{D_{(s)} (K,\pi)} $  is the largest value of the squared-momentum-transfer in the identified physical decay process.

It is worth remarking that in the SU$(4)$-flavour symmetry limit, $f_+^{D}(t)$ is the same as the elastic form factor for a charged pion-like meson constituted from a valence-quark and -antiquark with equal current masses \cite{Chen:2018rwz}.  Moreover, $f_-^{D}(t)\equiv 0$.  Hence, $f_-^{D}(t)$ should be a useful measure of SU$(4)$-flavour breaking.  Similarly,
$f_+^{D_s^d}/f_+^{D_d^s}$ and $f_+^{D_s^d}/f_+^{D_u^d}$ serve as gauges of SU$(3)$-flavour-symmetry breaking.  These features are correlated with the scalar form factor
\begin{equation}
\label{Eqf0}
f_0^D(t) = f_+^D(t) + \frac{t}{m_{D_{(s)}}^2 - m_{K,\pi}^2} f_-(t)\,,
\end{equation}
which measures the divergence of the transition current, $Q\cdot M^D(P,Q)$.

We note, too, that it is common to focus on the form factors $f_{+,0}(t)$ because each is separately characterised by a different resonance structure on $t \gtrsim t_m$: $f_{+}(t)$ connects with the vector meson $D^\ast_{(s)}$; and $f_{0}(t)$ with the analogous scalar resonance.  In contrast, $f_{-}(t)$ overlaps with both channels.  (These properties have been exemplified in studies of $K_{\ell 3}$ transitions \cite{Ji:2001pj, Chen:2012txa}.)

\subsection{Transition amplitudes}
We compute the matrix elements in Eqs.\,\eqref{EqMEs} at leading-order in a symmetry-preserving truncation scheme for the continuum bound-state equations \cite{Munczek:1994zz, Bender:1996bb}, \emph{i.e}.\ the rainbow-ladder (RL) truncation.  Focusing on $D_s^+\to K^0$, because the others are obvious by analogy:
\begin{align}
\nonumber
&_d M_\mu^{D^+_s}(P,Q)  = N_c {\rm tr}\int\frac{d^4 {\mathpzc s}}{(2\pi)^4}
\Gamma_{D_s}({\mathpzc s}+p/2;p) S_c({\mathpzc s}+p) \\
& \times i \Gamma_\mu^{cd}({\mathpzc s}+p,{\mathpzc s}-k) S_d({\mathpzc s}-k) \Gamma_K({\mathpzc s}-k/2;-k) S_s({\mathpzc s})\,,
\label{dMDs}
\end{align}
where the trace is over spinor indices and $N_c=3$.

There are three distinct types of matrix-valued functions in Eq.\,\eqref{dMDs}.  The simplest are the propagators for the dressed-quarks involved in the transition process: $S_f({\mathpzc s})$, $f=d,s,c$; then there are the Bethe-Salpeter amplitudes for the mesons involved: $\Gamma_{M}$, $M = D_s, K$; and, finally, the dressed vector piece of the $c\to d$ weak transition vertex: $\Gamma_\mu^{cd}$.  These functions are explained in Appendix~\ref{AppendixA}.  The scalar functions characterising the transition are obtained from Eq.\,\eqref{dMDs} using straightforward projections:
\begin{subequations}
\label{Projections}
\begin{align}
  f_+^{D_s^d}(t) & = \frac{t P_\mu - (m_{D_s}^2- m_K^2) Q_\mu}{t P^2 + (m_{D_s}^2- m_K^2)^2}
\, _d M_\mu^{D^+_s}(P,Q)\,, \\
 f_0^{D_s^d}(t) & = - \frac{Q_\mu}{m_{D_s}^2- m_K^2} \, _d M_\mu^{D^+_s}(P,Q)\,,
\end{align}
\end{subequations}
with $f_-^{D_s^d}(t)$ reconstructed via Eq.\,\eqref{Eqf0}.

\section{Computational Method and Results}
\label{SecCMR}
Predictions for the transition form factors can now be obtained by combining the quark propagators, Bethe-Salpeter amplitudes, and transition vertex, computed as described in Appendix\,\ref{AppendixA}, to form the integrand in Eq.\,\eqref{dMDs}; computing the integral as a function of $t$; and projecting the results according to Eq.\,\eqref{Projections}.  These steps can be completed in a straightforward manner so long as the difference between the current-masses of the quarks involved is not too large, \emph{e.g}.\ in the case of $K_{\ell 3}$ transitions \cite{Ji:2001pj}. (We verified this explicitly by repeating the analysis in Ref.\,\cite{Ji:2001pj}, obtaining consistent results for all calculated quantities.)  However, for both the $K$ and $\pi$ final states, owing to the analytic structure of the dressed-quark propagators and associated moving singularities in the complex-${\mathpzc s}^2$ domain sampled by the bound-state equations \cite{Maris:1997tm, Windisch:2016iud}, the direct approach fails when the current-mass of the heavier quark in the initial state exceeds $2.7$-times that of the $s$-quark.

Ref.\,\cite{Chang:2013nia} solved an analogous issue with pseudoscalar meson elastic electromagnetic form factors by using perturbation theory integral representations (PTIRs) \cite{Nakanishi:1969ph} for each matrix-valued function in the integrand defining the associated matrix element, thereby enabling a reliable computation of the form factor to arbitrarily large-$Q^2$.  However, constructing accurate PTIRs is time consuming;
and particularly so here because the complete set of integrands involves 46 distinct scalar functions, for each of which one would need to build a PTIR.

We therefore adopted a different approach.
%
%
Namely, we considered a fictitious pseudoscalar meson $P=D_{Qq}$, and computed its mass, $m_{P}$, and leptonic decay constant, $f_{P}$ as a function of $\hat m_Q$, the current-mass of the quark partnering the light-quark, $q$, in the initial state, up to a value $\hat m_Q = 2.7 \,\hat m_s$.  (Here $\hat m_f$ is the renormalisation point invariant current-mass for the $f$-quark.  Light quark values are listed in Eq.\,\eqref{musquark}.)
Then, using the Schlessinger point method (SPM) \cite{Schlessinger:1966zz, PhysRev.167.1411}, strengthened by the statistical sampling technique introduced in Refs.\,\cite{Chen:2018nsg, Binosi:2018rht, Binosi:2019ecz}, we built interpolations: $m_{P}(\hat m_Q)$, $f_{P}(\hat m_Q)$.

\begin{table}[t]
\caption{\label{Dstatic}
Computed values for static properties of mesons involved in the transitions studied herein, compared with empirical values \cite{Tanabashi:2018oca}.  Current-quark masses are given in Eqs.\,\eqref{mcquark}, \eqref{musquark}.
The results in Row~1 were obtained by direct computation.
The SPM was used to compute the values in Rows~3, 5.  In these cases, the uncertainty in our prediction expresses a $1\sigma$ confidence level on the SPM extrapolation, \emph{i.e}.\ 68\% of all SPM approximants give values that lie within the indicated band.
(All quantities in GeV.)
}
\begin{tabular}{l|cccc}\hline
  & $m_\pi$ & $f_\pi$ & $m_K$ & $f_K$\\\hline
herein & $0.135$ & $0.093$ & $0.494$ & $0.108$ \\
expt.\,\cite{Tanabashi:2018oca} & & $0.092$ & $0.494$ &  $0.110$ \\\hline
  & $m_{D}$ & $f_{D}$ & $m_{D_s}$ & $f_{D_s}$ \\\hline
herein & $1.86(7)$ & $0.150(5)$ & $1.95(4)$ & $0.188(8)$ \\
expt.\,\cite{Tanabashi:2018oca} & $1.87\phantom{(7)}$ & $0.153(7)$ & $1.97\phantom{(7)}$ & $0.177(3)$ \\\hline
  & $m_{D^\ast}$ & $m_{D_s^\ast}$ & $m_{S_{c \bar d}}$ & $m_{S_{c\bar s}}$\\\hline
herein & $2.11(5)$ & $2.15(4)$ &$2.12(3)$  & $2.25(4)$ \\
expt.\,\cite{Tanabashi:2018oca} & $2.01\phantom{(5)}$ & $2.11\phantom{(5)}$ & $2.30(2)$& $2.32\phantom{(5)}$ \\\hline
\end{tabular}
\end{table}

In explanation, the SPM is based on the Pad\'e approximant.  It can accurately reconstruct a function in the complex plane within a radius of convergence specified by that one of the function's branch points which lies nearest to the real domain from which the sample points are drawn.  Additionally, owing to the procedure's discrete nature and our statistical implementation, the reconstruction can also provide a reasonable continuation on a larger domain along with an estimate of the associated error.

At this point, capitalising on the strength of the statistical SPM, $\hat m_c$ could be determined by extrapolating the interpolating function, $m_{P}(\hat m_Q)$, and locating that value of the argument for which the projected meson mass matches the empirical value of the $D_{(s)}$ meson.  This exercise yielded
\begin{equation}
\label{mcquark}
\hat m_c = 1.93\,{\rm GeV};
\end{equation}
hence, $\hat m_c/\hat m_s = 12.0$ and, one-loop evolved to $\zeta_2=2\,$GeV, $m_c^{\zeta_2}=1.34\,$GeV.  These values are commensurate with those determined by other means \cite{Tanabashi:2018oca}.

To check consistency, we evaluated $f_{P}(\hat m_c)$ and compared with experiment.  The results are listed in Table~\ref{Dstatic}.  Evidently, the SPM delivers sound results for the masses and decay constants.  Confidence in the procedure is increased by noting that all values are consistent with those determined in Ref.\,\cite{Binosi:2018rht} by extrapolating in the other direction, \emph{viz}.\ from heavy to light current-masses.

Having determined the $c$-quark current-mass and validated the SPM in connection with static properties of $D_{(s)}$ mesons, we computed the $\hat m_Q$-dependence of $f_{+,0}^P(t)$ via direct calculation up to $\hat m_Q = 2.7\,\hat m_s$ and subsequently constructed SPM interpolations of $f_{+,0}^P(t;\hat m_Q)$, writing
\begin{subequations}
\label{mQfp0}
\begin{align}
f_+^P(t;\hat m_Q) & = \alpha_1(\hat m_Q) +   \frac{\alpha_2(\hat m_Q)}{1 - t/m_{V}^2}\,, \\
f_0^P(t;\hat m_Q) & = \alpha_1(\hat m_Q) + t \beta_2(\hat m_Q)  +   \frac{t^2 \beta_3(\hat m_Q)}{1 - t/m_{S}^2}\,.
\end{align}
\end{subequations}
These expressions capitalise on the fact that $f_+^P(0)=f_0^P(0)$, Eq.\,\eqref{Eqf0}, and exploit the known singularity structure of the weak vector transition vertex, so that $m_{V(S)}$ is the mass of the vector (scalar) state correlated with $P$.  These masses were calculated using Eq.\,\eqref{VectorVertex} in tandem with the SPM; and the results are listed in Table~\ref{Dstatic}.  The mean absolute relative error is 5(3)\%.

The coefficients $\alpha_{1,2}$, $\beta_{2,3}$ in Eq.\,\eqref{mQfp0} evolve with increasing $\hat m_Q$.  It is that behaviour we analysed using the SPM, a procedure which yielded the coefficients in Table~\ref{SPMcoefficients}.  Using these values, the formulae in Eqs.\,\eqref{mQfp0} deliver predictions for the $D_{s} \to K$ transition form factors.  Repeating the procedure, one also obtains the form factors describing $D\to\pi$, $D\to K$ transitions.

\begin{table}[t]
\caption{\label{SPMcoefficients}
Used in Eqs.\,\eqref{mQfp0}, these coefficients define predictions for all independent $D_{(s)}$ semileptonic transition form factors.
}
\begin{tabular}{l|cccc}\hline
  & $\alpha_1$ & $\alpha_2$ & $\beta_2/{\rm GeV}$ & $\beta_3/{\rm GeV}^2$\\\hline
$D_s\to K$ & $0.673(40)$ & $0.315(45)$ & $0.163(27)$ & $0.034(11)$ \\
$D\to \pi$ & $0.618(31)$ & $0.233(26)$ & $0.129(18)$ & $0.022(06)$ \\
$D\to K$ & $0.756(36)$ & $0.221(22)$ & $0.136(08)$ & $0.028(04)$ \\\hline
\end{tabular}
\end{table}

\section{Comparisons and Insights}
\label{SecCI}
\subsection{Form Factors}
\label{SecFF}
\begin{figure}[t]
\centerline{%
\includegraphics[clip, width=0.44\textwidth]{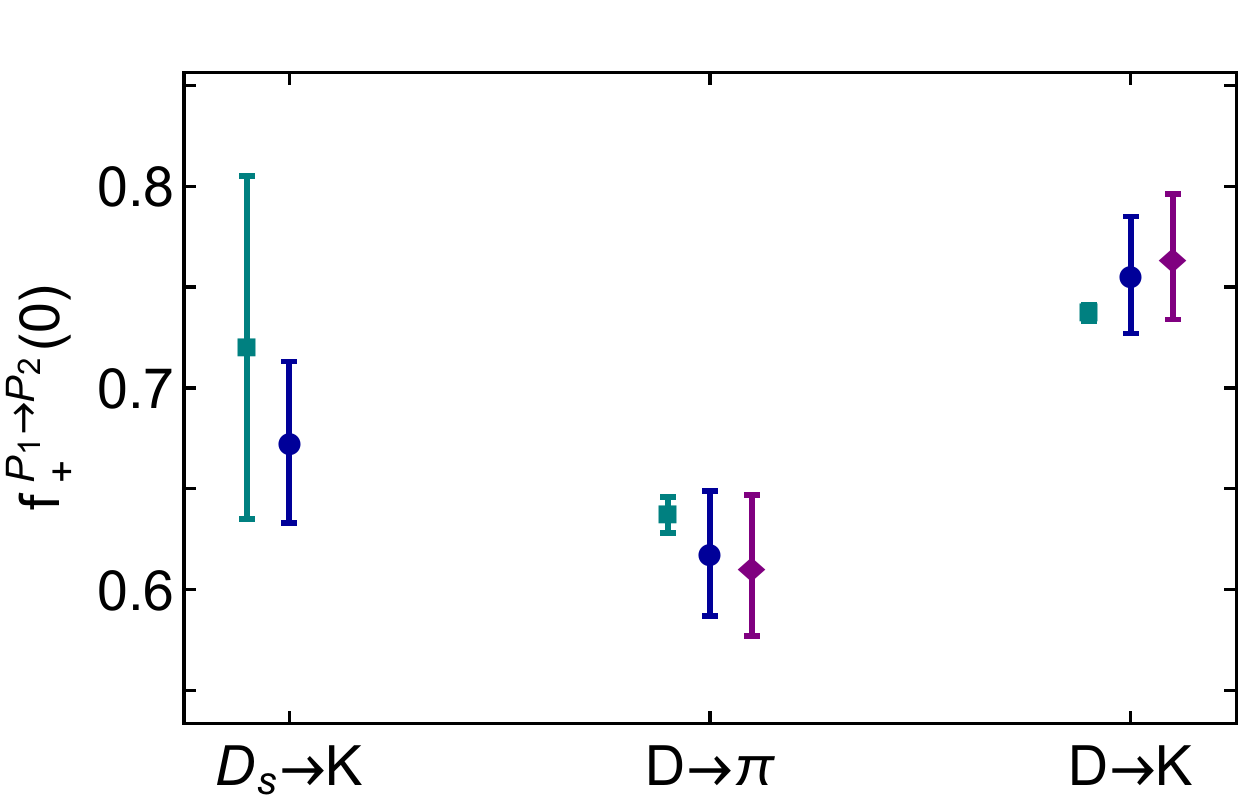}}
\caption{\label{figf0}
Pictorial representation of Table~\ref{fp0val}; namely, maximum recoil $(t=0)$ value of $D_{(s)}$ semileptonic transition form factors computed herein (blue circles) compared with inferences from experiment (cyan squares)  \cite{Ablikim:2015ixa, Ablikim:2018upe} and lQCD (purple diamonds) \cite{Lubicz:2017syv} (where the latter are available).
}
\end{figure}

Data is now available for $D_{(s)}$ semileptonic transition form factors \cite{Ablikim:2015ixa, Ablikim:2018upe}; and in Fig.\,\ref{figf0} and Table~\ref{fp0val} we compare our predictions for $f_{+,0}(t=0)$ with experiment and available lQCD results.  No parameters were varied in order to obtain our results and the agreement with both experiment and lQCD is good.  This is particularly important for $f_+^{D_s^d}(0)$ because no lQCD results are yet available and our result confirms the only available experiment \cite{Ablikim:2018upe}.

\begin{table}[t]
\caption{\label{fp0val}
Maximum recoil $(t=0)$ value of $D_{(s)}$ semileptonic transition form factors compared with inferences from experiment \cite{Ablikim:2015ixa, Ablikim:2018upe} and lQCD \cite{Lubicz:2017syv} (where available).  In each case, we also list our predictions for $f_-(t=0)$.
}
\begin{tabular}{l|ccc||c}\hline
  $f_+^{P_1 \to P_2}(0)$ & herein & expt. & lQCD & $-f_-^{P_1 \to P_2}(0)$ \\\hline
$D_s\to K$ & $0.673(40)$ & $0.720(85)$  &   & 0.553(65)\\
$D\to \pi$ & $0.618(31)$ & $0.637(09)$ & $0.612(35)$  & 0.362(28)\\
$D\to K$ & $0.756(36)$ & $0.737(04)$ & $0.765(31)$ & $0.277(45)$  \\\hline
\end{tabular}
\end{table}

\begin{figure}[t]
\centerline{%
\includegraphics[clip, width=0.44\textwidth]{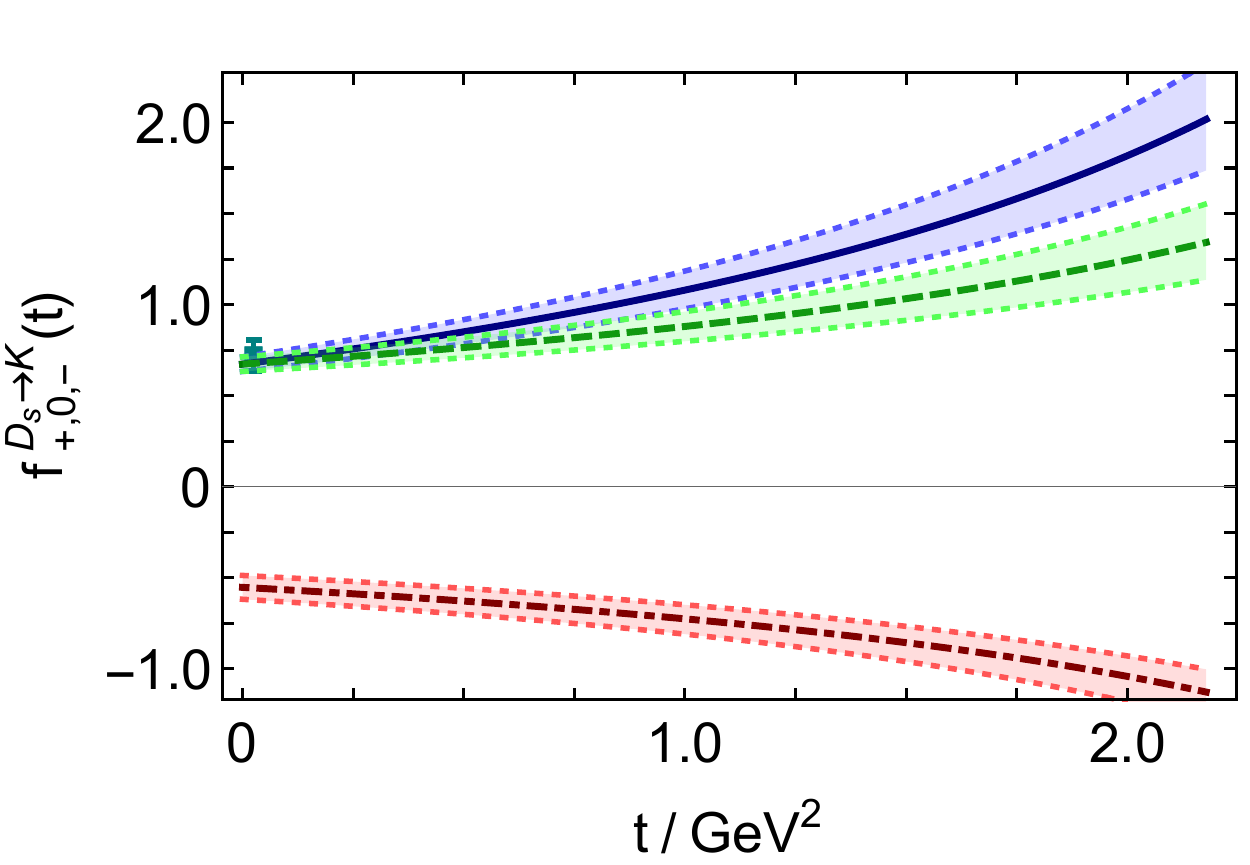}}
\caption{\label{FigDsK}
$D_s \to K$ semileptonic transition form factors, defined by Eqs.\,\eqref{Eqf0}, \eqref{mQfp0} with the associated coefficients listed in Table~\ref{SPMcoefficients}.
Legend: $f_+^{D_s^d}$ -- solid blue curve; $f_0^{D_s^d}$ -- dashed green curve; and $f_-^{D_s^d}$ -- dot-dashed red curve.  The shaded bands indicate the $1\sigma$ confidence level for the SPM extrapolations, \emph{i.e}.\ 68\% of all SPM approximants lie within the band centred on a given curve.
Empirical datum -- cyan square \cite{Ablikim:2018upe}.
}
\end{figure}

We draw our predictions for the $D_s \to K$ semileptonic transition form factors in Fig.\,\ref{FigDsK}.  Apart from the \mbox{$t=0$} datum in Table~\ref{fp0val} \cite{Ablikim:2018upe}, there are neither empirical data nor lQCD results for any of these three form factors.

\begin{figure}[t]
\includegraphics[clip, width=0.44\textwidth]{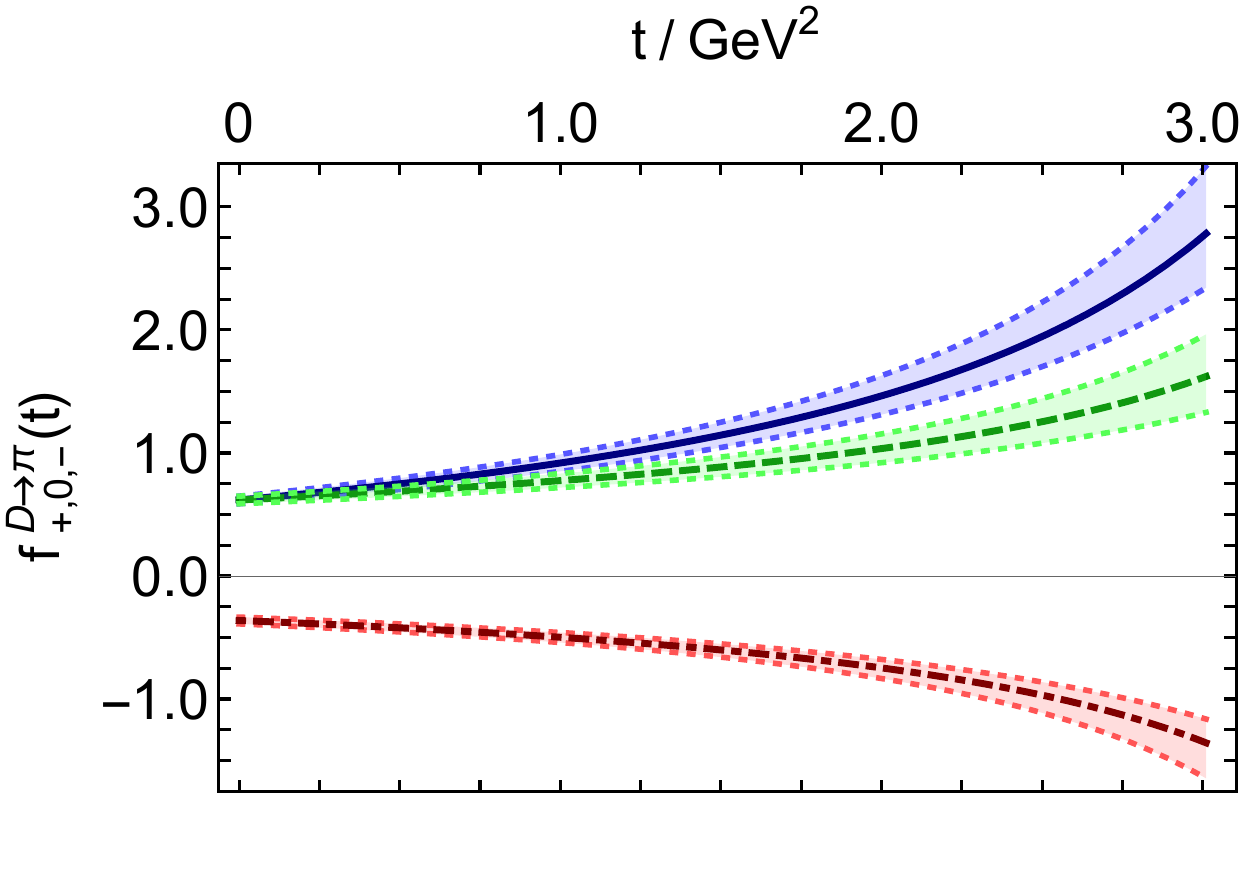}\vspace*{-5ex}

\includegraphics[clip, width=0.44\textwidth]{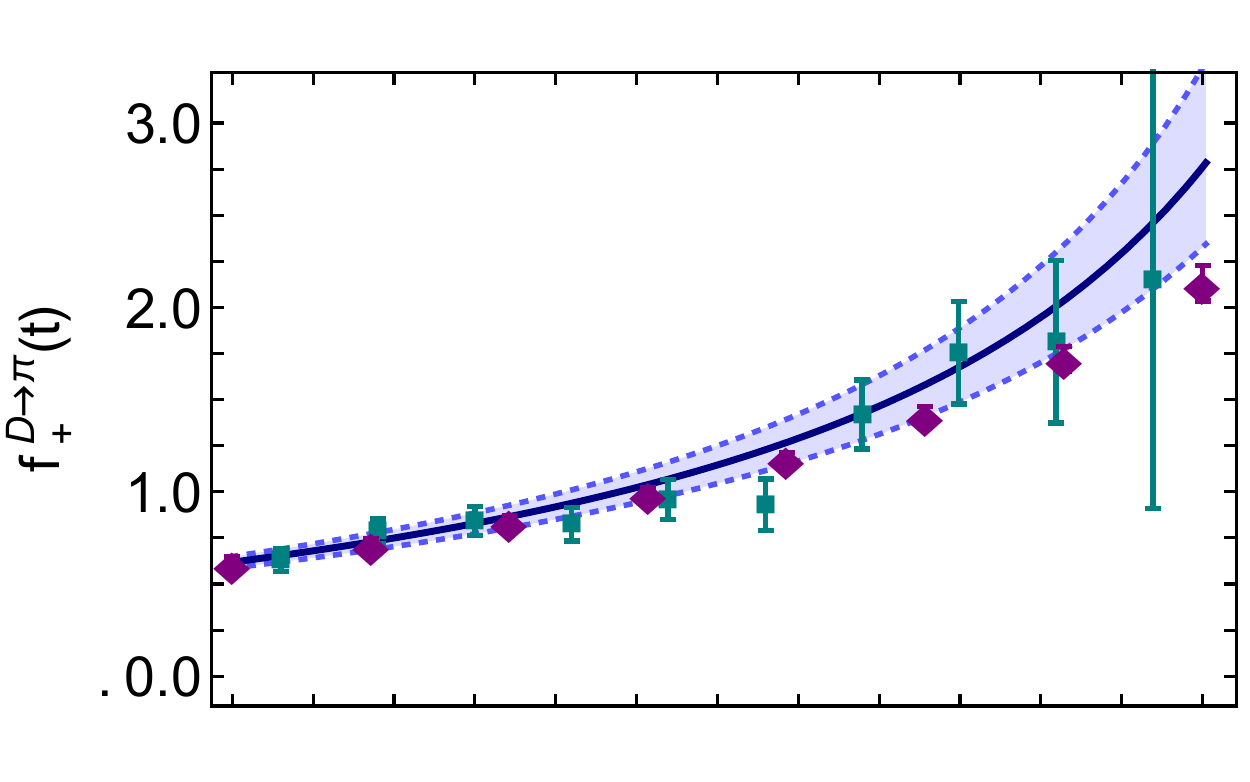}\vspace*{-5ex}

\includegraphics[clip, width=0.44\textwidth]{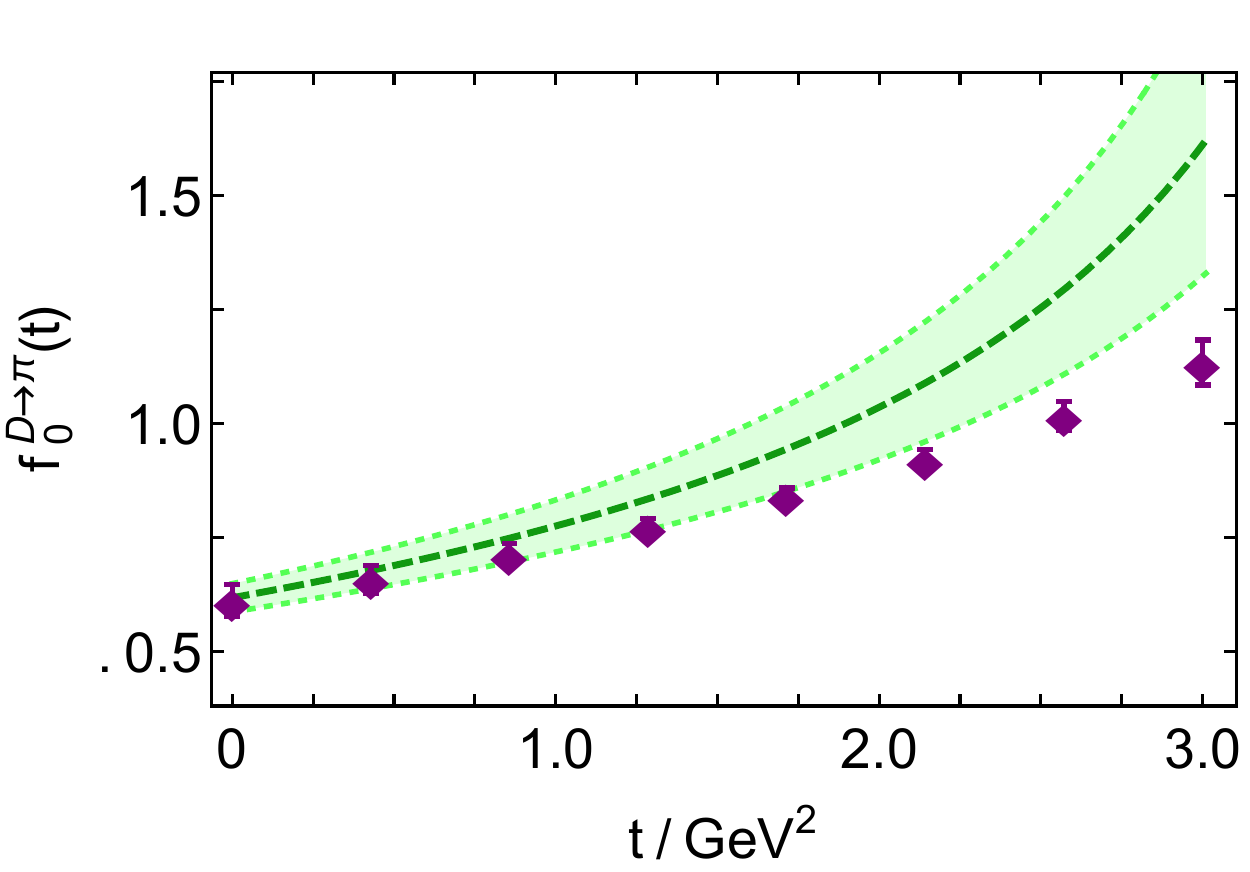}
\caption{\label{FigDpi}
$D \to \pi$ semileptonic transition form factors, defined by Eqs.\,\eqref{Eqf0}, \eqref{mQfp0} with the associated coefficients listed in Table~\ref{SPMcoefficients}.
Legend: $f_+^{D_u^d}$ -- solid blue curve; $f_0^{D_u^d}$ -- dashed green curve; and $f_-^{D_u^d}$ -- dot-dashed red curve.  The shaded bands indicate the $1\sigma$ confidence level for the SPM extrapolations. 
Empirical data -- cyan squares \cite{Ablikim:2015ixa}; and lQCD results -- purple diamonds \cite{Lubicz:2017syv}.
}
\end{figure}

Our calculated $D \to \pi$ semileptonic transition form factors are plotted in Fig.\,\ref{FigDpi}.  Referring to the middle panel, our result for $f_+^{D_u^d}(t)$ agrees with existing experiment \cite{Ablikim:2015ixa}.  On the other hand, the lQCD points lie systematically below our curves.  Turning to the bottom panel, we note that no data are available for $f_0^{D_u^d}(t)$; and here, too, the lQCD points lie systematically below our results.

\begin{figure}[t]
\includegraphics[clip, width=0.44\textwidth]{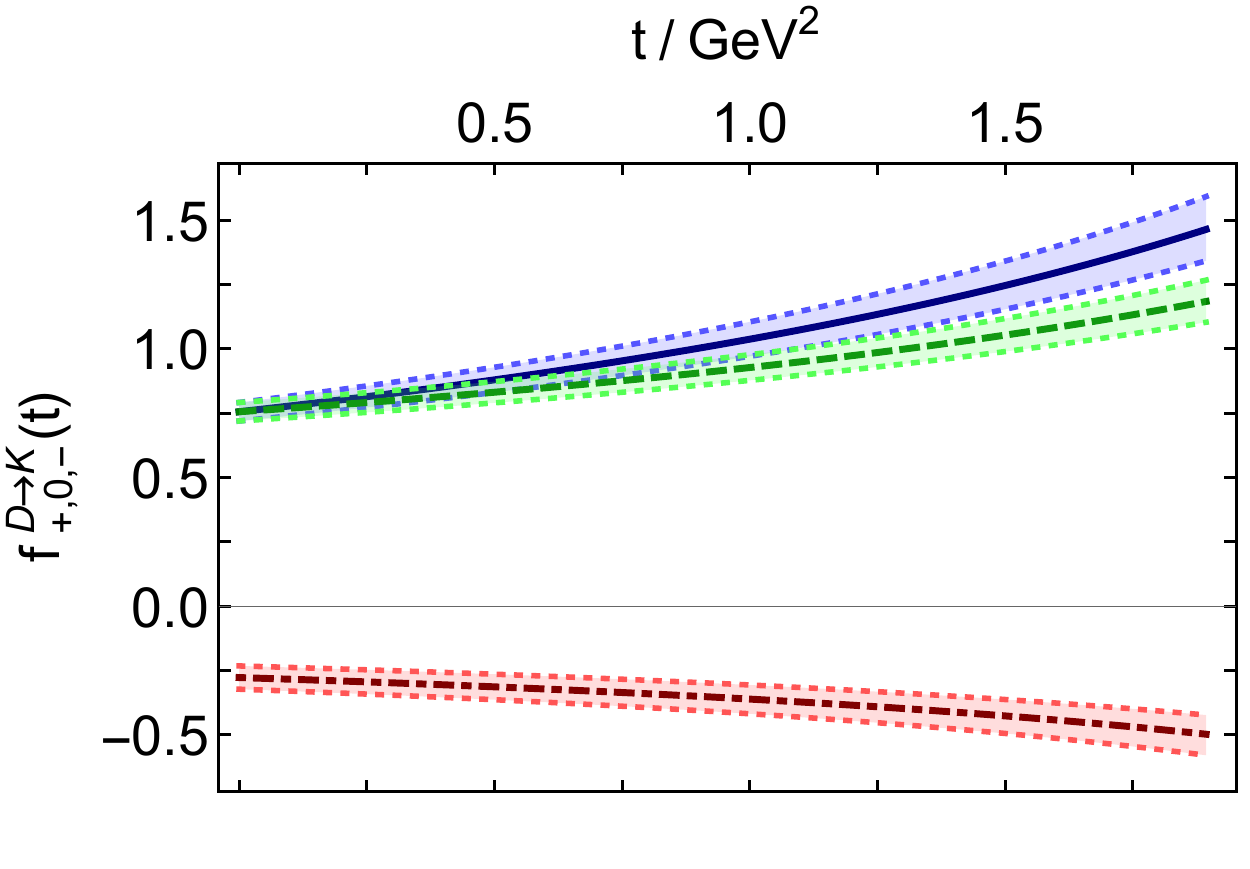}\vspace*{-5ex}

\includegraphics[clip, width=0.44\textwidth]{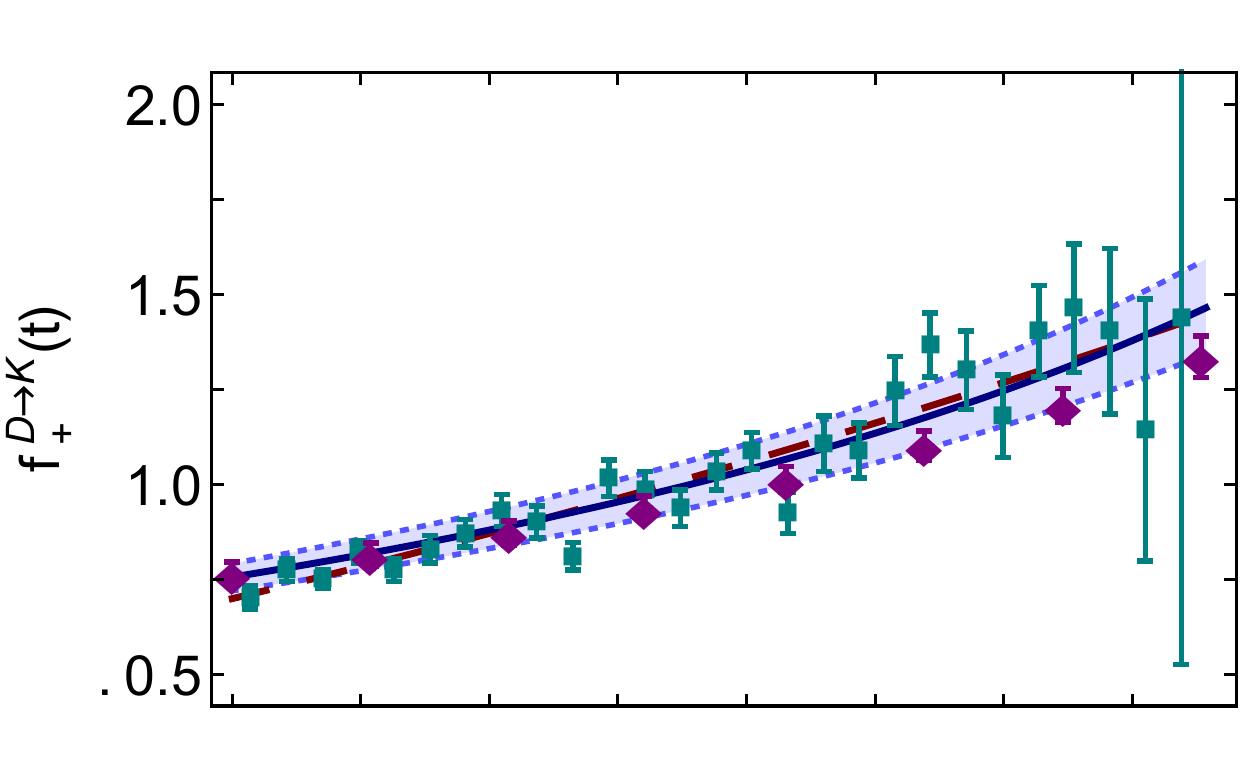}\vspace*{-5ex}

\includegraphics[clip, width=0.44\textwidth]{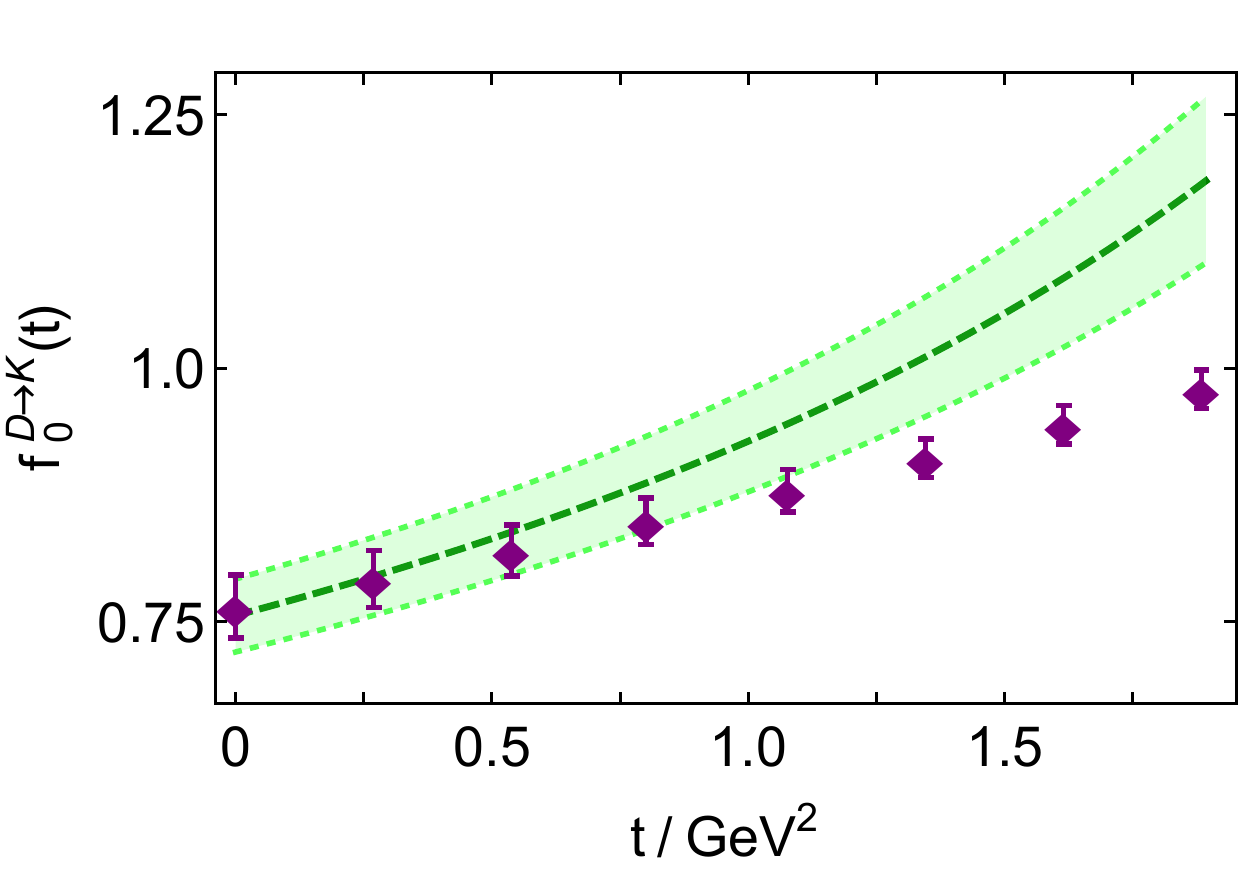}
\caption{\label{FigDK}
$D \to K$ semileptonic transition form factors, defined by Eqs.\,\eqref{Eqf0}, \eqref{mQfp0} with the associated coefficients listed in Table~\ref{SPMcoefficients}.
Legend: $f_+^{D_d^s}$ -- solid blue curve; $f_0^{D_d^s}$ -- dashed green curve; and $f_-^{D_d^s}$ -- dot-dashed red curve.  The shaded bands indicate the $1\sigma$ confidence level for the SPM extrapolations. 
Empirical data -- cyan squares \cite{Ablikim:2015ixa}; and lQCD results -- purple diamonds \cite{Lubicz:2017syv}.
Long-dashed purple curve: least-squares fit to data -- $f_+^{\rm fit}(t) = (0.70 + 0.27 t)/(1-0.089 t)$.
}
\end{figure}

We plot our calculated $D \to K$ semileptonic transition form factors in Fig.\,\ref{FigDK}.  $f_+^{D_d^s}(t)$ (middle panel) agrees fairly well with experiment \cite{Ablikim:2015ixa}; albeit comparison with the simple least-squares fit to data indicates that it may be a little too large at small $t$.  The lQCD points typically lie at the lower edge of our range for $f_+^{D_d^s}(t)$.  Regarding $f_0^{D_d^s}(t)$ (bottom panel), again the lQCD points typically lie below our result.

\subsection{Branching Fractions}
\label{SecBF}
With computed transition form factors and available experimental data, one can place constraints on the CKM matrix elements $|V_{cd}|$ and $|V_{cs}|$.
Here, the $D_s\to K$ transition is most interesting because: the experimental uncertainty is largest; lQCD results for $f_{+,0,-}^{D_s^d}(t)$ are not yet available; and we have predictions for these form factors.

The partial width for the $D_s^+ \to K^0 e^+ \nu_e$ transition is given by \cite{Bajc:1995km}:
\begin{subequations}
\label{DecayFraction}
\begin{align}
\Gamma_{D_s K} & = \frac{G_F^2 |V_{cd}|^2}{24 \pi^3}
\int_0^{y_m^{D_s K}} \!\! dy\,[f_+^{D_s^d}(y \, m_{D_s}^2) ]^2 k_{D_s K}^3(y)\,,\\
k_{D_s K}^2(t) & = (m_{D_s}^2 (1-y) + m_K^2)^2/[4 m_{D_s}^2] - m_K^2,
\end{align}
\end{subequations}
with $G_F = 1.166 \times 10^{-5}\,$GeV$^{-2}$.  Using our result for $f_+^{D_s^d}(t)$, the associated branching fraction is
\begin{align}
\label{BFDsK}
{\mathpzc B}_{D_s^+ \to K^0 e^+ \nu_e} = (0.264(13) |V_{cd}|)^2\,.
\end{align}
Combining Eq.\,\eqref{BFDsK} with the branching fraction reported in Ref.\,\cite{Ablikim:2018upe}: $3.25(38) \times 10^{-3}$, one finds
\begin{equation}
\label{Vcd1}
|V_{cd}| = 0.216(17)\,,
\end{equation}
a result that is consistent with the average in Ref.\,\cite{Tanabashi:2018oca}: $|V_{cd}| = 0.218(4)$.  Alternatively, using this average value, Eqs.\,\eqref{DecayFraction} yield:
\begin{align}
\label{BFDsKE}
{\mathpzc B}_{D_s^+ \to K^0 e^+ \nu_e} &= 3.31(33) \times 10^{-3}.
\end{align}
To obtain something new from $D_s\to K$ transitions, the precision of both experiment and theory must improve.

We collect our results for branching fractions in Table~\ref{Branching}.

\begin{table}[t]
\caption{\label{Branching}
Computed branching fractions (Row~1) compared with empirical results (Row~2) drawn from Refs.\,\cite{Ablikim:2015ixa, Ablikim:2018upe}: in these rows, each entry should be multiplied by $10^{-3}$.  Row~3 -- value of $|V_{c(d,s)}|$ required to reproduce Row~2 using our results for $f_+^{D_{(s)}(\pi,K)}$.
Ref.\,\cite{Ablikim:2015ixa} lists $|V_{cd}| = 0.216(10)$, $|V_{cs}|= 0.960(25)$; and
Ref.\,\cite{Tanabashi:2018oca}: $|V_{cd}| = 0.218(04)$, $|V_{cs}|= 0.997(17)$.
}
\begin{tabular}{l|cc|c}\hline
    & ${\mathpzc B}_{D_s^+ \to K^0 e^+ \nu_e}$ & ${\mathpzc B}_{D^0 \to \pi^- e^+ \nu_e}$ & ${\mathpzc B}_{D^0 \to K^- e^+ \nu_e}$ \\\hline
herein & $3.31(33)\phantom{7}$ & $2.73(22)\phantom{7} $ & $38.34(2.82) $ \\
expt.\,\cite{Ablikim:2015ixa, Ablikim:2018upe} &
$3.25(38)\phantom{7}$ & $2.95(05)\phantom{7} $ & $35.05(0.36)$\\\hline
herein & $0.216(17)$ & $0.227(10)$ & $0.953(34)$\\\hline
\end{tabular}
\end{table}

Analogies of Eq.\,\eqref{DecayFraction} can be used for $D^0 \to (\pi,K)$ transitions; and with our result for $f_+^{D_u^d}(t)$:
\begin{align}
\label{BFDpi}
{\mathpzc B}_{D^0 \to \pi^- e^+ \nu_e} = (0.240(10) |V_{cd}|)^2\,.
\end{align}
Combining Eq.\,\eqref{BFDpi} with the branching fraction reported in Ref.\,\cite{Ablikim:2015ixa}: $2.95(05) \times 10^{-3}$, one finds
\begin{equation}
\label{Vcd2}
|V_{cd}| = 0.227(10)\,,
\end{equation}
consistent with Eq.\,\eqref{Vcd1}.  On the other hand, with $|V_{cd}| = 0.218(4)$:
\begin{align}
\label{BFDpiE}
{\mathpzc B}_{D^0 \to \pi^- e^+ \nu_e} &= 2.73(22) \times 10^{-3}.
\end{align}

An average of Eqs.\,\eqref{Vcd1}, \eqref{Vcd2} yields $|V_{cd}|=0.221(9)$.

Considering $D^0\to K^- e^+ \nu_e$, our result for $f_+^{D_d^s}(t)$ produces
\begin{equation}
{\mathpzc B}_{D^0 \to K^- e^+ \nu_e} = (0.196(7) |V_{cs}|)^2;
\end{equation}
hence, with $|V_{cs}|= 0.997(17)$ \cite{Tanabashi:2018oca} one obtains
\begin{equation}
{\mathpzc B}_{D^0 \to K^- e^+ \nu_e} = 3.83(28) \times 10^{-2}.
\end{equation}
This may be compared with the empirical value reported in Ref.\,\cite{Ablikim:2015ixa}: ${\mathpzc B}_{D^0 \to K^- e^+ \nu_e}=3.505(36) \times 10^{-2}$.  Agreement with this fraction would require:
\begin{equation}
|V_{cs}| = 0.953(34)\,.
\end{equation}

These comparisons suggest that our result for $f_+^{D_d^s}(t)$ may be \emph{marginally} too large on $t\simeq 0$.  To explore this possibility, we repeated the analysis using the simple fit to experimental data depicted as the dashed purple curve in Fig.\,\ref{FigDK} -- middle panel, retaining the uncertainty of our calculated result, and obtained
\begin{subequations}
\begin{align}
{\mathpzc B}_{D^0 \to K^- e^+ \nu_e} & = (0.194(7) |V_{cs}|)^2,\\
 & \stackrel{|V_{cs}|=0.997(17)}{=} 3.73(27) \times 10^{-2},\\
\mbox{\rm or} \; |V_{cs}| & \stackrel{{\mathpzc B}_{D^0 \to K^-} \mbox{\footnotesize in~Ref.\,\cite{Ablikim:2015ixa}}}{=} 0.966(35)\,.
\end{align}
\end{subequations}
Evidently, this replacement achieves no material improvement, but the test does confirm consistency of our results with the analysis in Ref.\,\cite{Ablikim:2015ixa}.


\subsection{Flavour-Symmetry Breaking}
\label{SecFSB}
Predictions for the collection of $D_{(s)}$ semileptonic transition form factors also enable examination of the interplay between EHM and Higgs-related mass generation in QCD's matter sector.  For example, given that $m_{D_s} \approx m_D$, then windows on SU$(3)$-flavour symmetry-breaking are provided by the ratio of associated leptonic decay constants and aspects of $D_{(s)}\to K$ transitions.  With this in mind, consider Table~\ref{Dstatic}:
\begin{equation}
\label{ldcratio}
\frac{f_{D_s}}{f_D} = 1.25(7) \approx \frac{f_K}{f_\pi}= 1.16 \; (1.20_{\rm expt.})\,;
\end{equation}
and Table~\ref{fp0val}:
\begin{equation}
\label{sltfratio}
\frac{f_+^{D_d^s}(0)}{f_+^{D_s^d}(0)} = 1.12(09) \approx \frac{f_K}{f_\pi}\,.
\end{equation}

\begin{figure}[t]
\centerline{%
\includegraphics[clip, width=0.44\textwidth]{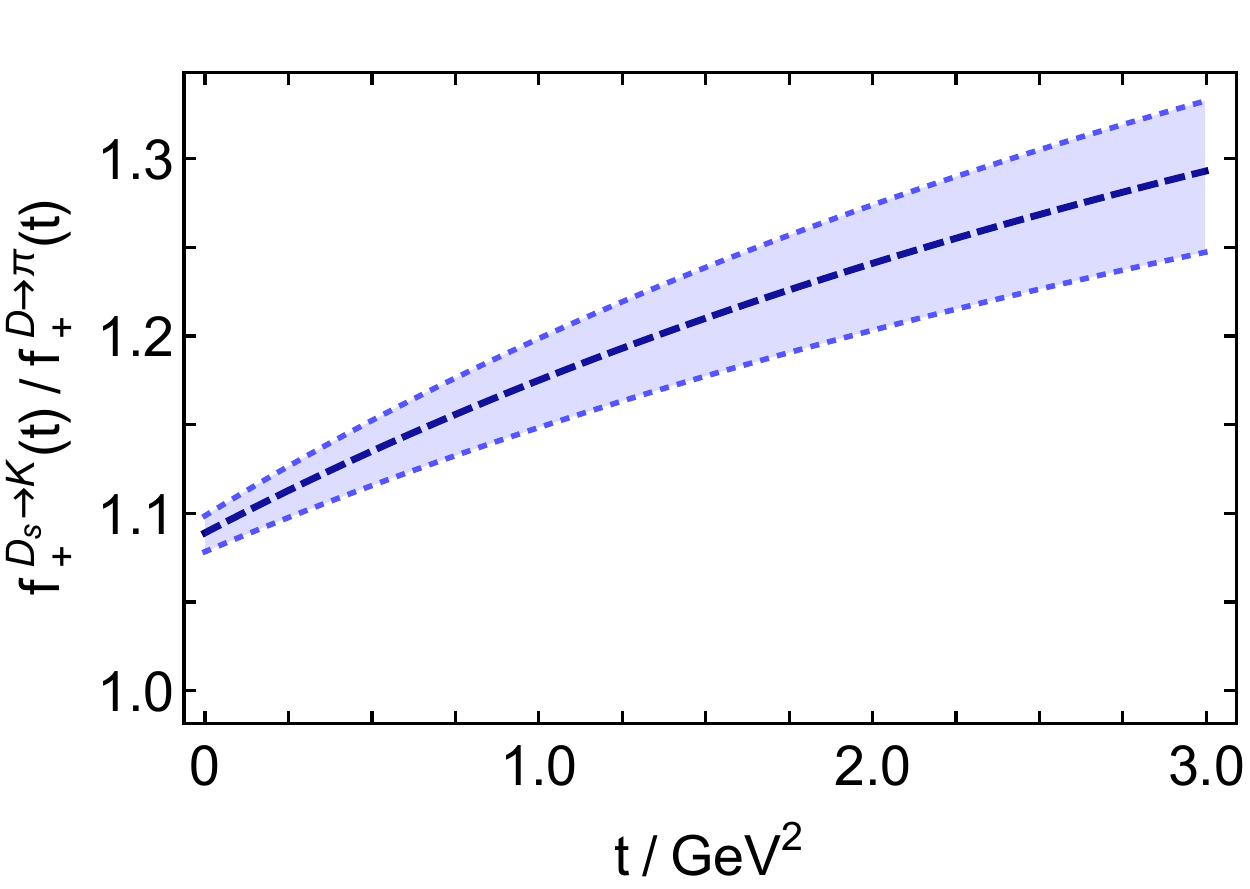}}
\caption{\label{USpin}
Computed $t$-dependence of the ratio in Eq.\,\eqref{USpinfratio}: the ``U-spin symmetry'' hypothesis becomes quantitatively unreliable as $t$ increases away from the maximum recoil point.
}
\end{figure}

The ratio in Eq.\,\eqref{sltfratio} simultaneously compares (\emph{i}) dynamical corrections to the $c\to s$ and $c\to d$ vertices and (\emph{ii}) processes with different interaction spectators: $\bar s$-quark \emph{cf}.\,$\bar u$.  A simpler quantity is
\begin{equation}
\label{USpinfratio}
\frac{f_+^{D_s^d}(0)}{f_+^{D_u^d}(0)} = 1.09(08) \,,
\end{equation}
where the result follows from Table~\ref{fp0val}.  For this ratio, the transition vertices involved are identical; only the spectators are different.  Plainly, there is little flavour sensitivity at $t=0$; but as revealed by Fig.\,\ref{USpin}, the ratio increases as $t$ ranges over the physical domain, undermining quantitative accuracy of the ``U-spin symmetry'' hypothesis in $D_{(s)}$ decays \cite{Gronau:2000zy}.  It is likely to work better for heavy+light pseudoscalars containing a $b$-quark \cite{Ivanov:2007cw}.

Plainly, the scale of SU$(3)$-flavour symmetry breaking is commensurate in all these systems.  Looking further, one finds that the ratios in Eqs.\,\eqref{ldcratio}, \eqref{sltfratio} are also similar in size to
the skewing of the kaon's leading-twist parton distribution amplitude (PDA) with respect to the asymptotic PDA profile \cite{Shi:2015esa}
and analogous distortions of the kaon's valence $u$- and $\bar s$-quark distribution functions \cite{Chen:2016sno}.
All these ratios are much smaller than
\begin{equation}
 \hat m_s/\hat m_d \approx 25\,.
 \end{equation}
This fact and the confluence of results highlighted above emphasises again that the observable magnitude of SU$(3)$-flavour symmetry-breaking in hadron properties is determined by EHM, which is directly expressed in the infrared value for the ratio of $s$ and $d$-quark mass functions \cite{Maris:1997tm, Shi:2015esa, Chen:2016sno}:
\begin{equation}
M_s^E/M_d^E=1.25(9),
\end{equation}
where the Euclidean constituent-quark mass, $M_f^E$, is a nonperturbative analogue of the so-called pole mass \cite{Tanabashi:2018oca}.

\section{Summary and Perspective}
\label{Epilogue}
We studied the leptonic and semileptonic decays of $D_{(s)}$ mesons using a well-constrained symmetry-preserving continuum treatment of the meson bound-state problem in quantum field theory, thereby unifying the treatment of these features of such systems with analogous properties of $\pi$ and $K$ mesons.

Our predictions for the $D_{(s)}$ transition form factors agree with available experimental data [Sec.\,\ref{SecFF}].  On the other hand, results obtained using lattice-regularised QCD typically lie below our results.
Additionally, our computed form factors deliver values for the $D_{(s)}\to (K,\pi) e^+\nu_e$ branching fractions which match those measured experimentally [Sec.\,\ref{SecBF}].
Subsequently, having calculated all $D_{(s)}$ transition form factors, we analysed the character of SU$(3)$-flavour symmetry-breaking; finding that, as in the $\pi$-$K$ sector, the observable magnitude of this effect is determined by the scales associated with emergent mass generation, not those originating with the Higgs mechanism [Sec.\,\ref{SecFSB}].

With the validity of our framework and computational algorithms supported by the results described herein, it is natural to extend this analysis to semileptonic $D_{(s)}\to V$ decays, where $V$ is a light-quark vector-meson, and also to the leptonic and semileptonic decays of $B$, $B_s$, $B_c$ mesons.  Such efforts are underway.  Kindred decays of baryons could also be treated on an equal footing using the Poincar\'e covariant Faddeev equation \cite{Eichmann:2011vu, Wang:2018kto, Qin:2019hgk} and the same symmetry-preserving truncations.

\begin{acknowledgments}
%
%
We are grateful for constructive comments from B.-L.~Li, Y.~Lu, Y.-Z.~Xu and Z.-N.~Xu.
Work supported by:
National Natural Science Foundation of China (under Grant Nos.\,11805097, 11847024, 11905107);
Jiangsu Provincial Natural Science Foundation of China (under Grant Nos.\,BK20180323, BK20190721);
Jiangsu Province \emph{Hundred Talents Plan for Professionals};
Innovation Program of Jiangsu Province;
Nanjing University of Posts and Telecommunications Science Foundation, under Grant No.\,NY129032;
and
Natural Science Foundation of the Jiangsu Higher Education Institutions of China, under Grant No.\,19KJB140016.
\end{acknowledgments}

\appendix
\section{Elements constituting the transition amplitude}
\label{AppendixA}
\subsection{Dressed-quark propagators}
In RL truncation, the gap equation for the dressed-propagator of a quark with bare mass $m_f^{\rm bm}(\Lambda)$ takes the form:
\begin{subequations}
\label{gendseN}
\begin{align}
S^{-1}(k) & = [i\gamma\cdot k + M_f(k^2)]/Z_f(k^2) \\
%
& = Z_2 \,(i\gamma\cdot k + m_f^{\rm bm}) + \Sigma_f(k)\,,\\
\label{EqSigma}
\Sigma_f(k)& =  \int^\Lambda_{d{\mathpzc s}} \!\! {\mathpzc G}_{\mu\nu}(k-{\mathpzc s}) \frac{\lambda^a}{2}\gamma_\mu S_f({\mathpzc s}) \frac{\lambda^a}{2}\gamma_\nu \,,
\end{align}
\end{subequations}
where:
$Z_2$ is the quark wave-function renormalisation constant, with $\zeta$ the renormalisation point; and
$\int^\Lambda_{d{\mathpzc s}}$ represents a Poincar\'e invariant regularisation of the four-dimensional Euclidean integral, with $\Lambda$ the regularisation mass-scale. (A Pauli-Villars-like scheme is usually adequate \cite{Holl:2005vu} and  renormalisation is performed in the chiral limit so that $Z_2$ is flavour-independent \cite{Chang:2008ec}.)  Following Ref.\,\cite{Maris:1997tm}, we choose $\zeta=19\,$GeV$=:\zeta_{19}$: physical quantities do not depend on the value of $\zeta$.

In Eq.\,\eqref{dMDs}, ${\mathpzc G}_{\mu\nu}$ is the quark-quark interaction appropriate for RL truncation, which is explained in Ref.\,\cite{Qin:2011dd, Qin:2011xq}:
\begin{equation}
\label{KDinteraction}
 {\mathpzc G}_{\mu\nu}(k)  = \tilde{\mathpzc G}(k^2) T_{\mu\nu}(k)\,,
\end{equation}
with $k^2 T_{\mu\nu}(k) = k^2 \delta_{\mu\nu} - k_\mu k_\nu$ and ($u=k^2$)
\begin{align}
\label{defcalG}
 \tfrac{1}{Z_2^2}\tilde{\mathpzc G}(u) & =
 \frac{8\pi^2 D}{\omega^4} e^{-u/\omega^2} + \frac{8\pi^2 \gamma_m \mathcal{F}(u)}{\ln\big[ \tau+(1+u/\Lambda_{\rm QCD}^2)^2 \big]}\,,
\end{align}
where $\gamma_m=4/\beta_0$, $\beta_0=11 - (2/3)n_f$, $n_f=4$,
$\Lambda_{\rm QCD}=0.234\,$GeV,
$\tau={\rm e}^2-1$ $(\ln {\rm e} = 1)$,
and ${\cal F}(u) = \{1 - \exp(-u/[4 m_t^2])\}/u$, $m_t=0.5\,$GeV.
The evolution of Eqs.\,\eqref{KDinteraction}, \eqref{defcalG} is reviewed in Ref.\,\cite{Qin:2011dd} and their relation to QCD is elaborated in Ref.\,\cite{Binosi:2014aea}.  Here we note only that the interaction is (\emph{a}) deliberately consistent with that obtained in studies of QCD's gauge sector and (\emph{b}) preserves QCD's one-loop renormalisation group behaviour.

Experience has shown \cite{Eichmann:2008ef, Qin:2011dd, Qin:2011xq, Eichmann:2012zz, Binosi:2014aea, Wang:2018kto, Chen:2018rwz, Qin:2019hgk} that Eq.\,\eqref{defcalG} is a one-parameter \emph{Ansatz} because observable properties of light-quark ground-state vector- and flavour-nonsinglet pseudoscalar-mesons are practically insensitive to variations of $\omega \in [0.4,0.6]\,$GeV so long as
\begin{equation}
 \varsigma^3 := D\omega = {\rm constant}.
\label{Dwconstant}
\end{equation}
The value of $\varsigma$ is usually chosen to reproduce the measured value of the pion's leptonic decay constant, $f_\pi$.  In RL truncation this requires
\begin{equation}
\label{varsigmalight}
\varsigma  =0.80\,{\rm GeV.}
\end{equation}
We employ $\omega=0.5\,$GeV, the midpoint of the domain of insensitivity.

\subsection{Bethe-Salpeter amplitudes}
The RL Bethe-Salpeter equation for a pseudoscalar meson, $P$, constituted from a valence $f$-quark and a valence $g$-antiquark is:
\begin{align}
\Gamma_{P}^{f\bar g}&(k;Q) = \int\frac{d^4 {\mathpzc s}}{(2\pi)^4}
 {\mathpzc G}_{\mu\nu}(k-{\mathpzc s})  \nonumber\\
  & \quad \times \frac{\lambda^a}{2}i\gamma_\mu
 S_f({\mathpzc s}_+) \Gamma_{P}^{f\bar g}({\mathpzc s};Q)
 S_g({\mathpzc s}_-) \frac{\lambda^a}{2} i\gamma_\nu \,,
\end{align}
where ${\mathpzc s}_+ = {\mathpzc s} + \eta Q $, ${\mathpzc s}_- = {\mathpzc s} - (1-\eta) Q $ and the quark propagators must be computed using Eq.\,\eqref{gendseN}.  The solution has the form $(\bar{k} = [k_++k_-]/2)$:
\begin{align}
\Gamma_{P}^{f\bar g}& (k;Q) = i \gamma_5\left[ E_{P}^{f\bar g}(\bar k;Q) + \gamma\cdot Q F_{P}^{f\bar g}(\bar k;Q) \right. \nonumber \\
& \quad \left. + \gamma\cdot \bar k G_{P}^{f\bar g}(\bar k;Q) + \sigma_{\mu\nu} \bar k_\mu Q_\nu H_{P}^{f\bar g}(\bar k;Q) \right]\,. \label{BSEamp}
\end{align}
In a symmetry-preserving framework, no measurable quantity is sensitive to the value of $\eta \in [0,1]$, \emph{i.e}.\ to the definition of relative momentum within the bound state \cite{Maris:1997tm}.  The choice we make is convenient because it ensures that the scalar functions in Eq.\,\eqref{BSEamp} are even under $\bar k \cdot Q \to - \bar k \cdot Q$.

The leptonic decay constant for this pseudoscalar meson, $f_P$, is obtained from the following expression:
\begin{subequations}
\begin{align}
\label{fPequation}
f_P Q_\mu & = Z_2 N_c {\rm tr}_{\rm D} \int^\Lambda_{dk} \!\!\gamma_5\gamma_\mu \chi_P^{f\bar g}(k;Q) \,,\\
\chi_P^{f\bar g}(k;Q) & = S_f(k_+) \Gamma_{P}^{f\bar g}(k;Q)  S_g(k_-) \,,
\end{align}
\end{subequations}
where the trace is over spinor indices.  Naturally, the integral in Eq.\,\eqref{fPequation} must be defined in the same manner as that in Eq.\,\eqref{gendseN}, using the same renormalisation point and regularisation scale.

With the following choices for the renormalisation group invariants (in GeV):
\begin{equation}
\label{musquark}
\hat m_{u=d} = 0.0068\,, \; \hat m_s = 0.162\,,
\end{equation}
which correspond to $m_u^{\zeta_{19}} = 0.0034\,$GeV, $m_s^{\zeta_{19}} = 0.082\,$GeV and one-loop evolved values ($\zeta_2=2\,$GeV),
\begin{equation}
m_u^{\zeta_{2}} = 0.0047,\;  m_s^{\zeta_{2}} = 0.112\,,
\end{equation}
one obtains the masses and decay constants in Table~\ref{Dstatic}.  These values of the light-quark current-masses are commensurate with those obtained via other means \cite{Tanabashi:2018oca}.

\subsection{Weak vector transition vertex}
The vector component of the $c\to d$ weak transition vertex is computed from the following inhomogeneous Bethe-Salpeter equation:
\begin{align}
&\Gamma_\rho^{cd}(p,k) = Z_2 \gamma_\rho
+ \int^\Lambda_{d{\mathpzc s}} \!\!
 {\mathpzc G}_{\mu\nu}({\mathpzc s}) \frac{\lambda^a}{2}i\gamma_\mu  \nonumber\\
  & \quad \times
 S_c({\mathpzc s}+p) \Gamma_\rho^{cd}({\mathpzc s}+p,{\mathpzc s}-k)
 S_d({\mathpzc s}-k) \frac{\lambda^a}{2} i\gamma_\nu \,.
 \label{VectorVertex}
\end{align}
This vertex satisfies a Ward-Green-Takahashi identity \cite{Ward:1950xp, Green:1953te, Takahashi:1957xn}
\begin{align}
\label{WGTI}
(p-k)_{\rho} i {\Gamma}^{cd}_{\mu}(p,k;\zeta) & = S_c^{-1}(p;\zeta) - S_d^{-1}(k;\zeta) \nonumber \\
 & \quad - (m_c^\zeta-m_d^\zeta)\Gamma^{cd}_I(p,k;\zeta)\,,
%
\end{align}
where $\Gamma_I^{cd}$ is an analogous Dirac-scalar vertex.  (The axial-vector piece of the weak transition vertex cannot contribute to a $0^- \to 0^-$ transition in the Standard Model.)  We have here made the renormalisation scale explicit, to mark the character of the current-quark masses.

When considering the electromagnetic current, $f\to f$, the solution of the analogous equation involves 11 independent terms, each with its own scalar coefficient function \cite{Ball:1980ay}: owing to the analogue of Eq.\,\eqref{WGTI}, three of these are determined by the dressed-quark propagator, leaving eight coupled equations to solve \cite{Maris:1999bh}.

In the present case, however, the active presence of $\Gamma^{cd}_I$ in Eq.\,\eqref{WGTI} entails that the vertex dynamics is not purely transverse; hence, the solution for $\Gamma_\rho^{cd}$ involves 12 independent scalar functions to be obtained from associated, coupled integral equations.  This task can readily be accomplished by separating the vertex into transverse and longitudinal components, choosing Dirac-matrix bases for both which are free of kinematic singularities \cite{Qin:2013mta}; and solving the resulting integral equations using now well-known algorithms \cite{Maris:1997tm, Krassnigg:2009gd}.

It is worth reiterating that the transverse part of $\Gamma_\rho^{cd}(p,k)$ exhibits a singularity when $(p-k)^2$ enters the neighbourhood of the mass of the $D^\ast$ meson.  The same is true for the longitudinal part in the neighbourhood of the mass of the analogous scalar meson.  In RL truncation, both singularities are simple poles.


\end{document}